# A Tour of Unsupervised Deep Learning for Medical Image Analysis

**Khalid Raza\* and Nripendra Kumar Singh**
Department of Computer Science, Jamia Millia Islamia, New Delhi
kraza@jmi.ac.in

**December 13, 2018**

**Abstract**

Interpretation of medical images for diagnosis and treatment of complex disease from high-dimensional and heterogeneous data remains a key challenge in transforming healthcare. In the last few years, both supervised and unsupervised deep learning achieved promising results in the area of medical imaging and image analysis. Unlike supervised learning which is biased towards how it is being supervised and manual efforts to create class label for the algorithm, unsupervised learning derive insights directly from the data itself, group the data and help to make data driven decisions without any external bias. This review systematically presents various unsupervised models applied to medical image analysis, including autoencoders and its several variants, Restricted Boltzmann machines, Deep belief networks, Deep Boltzmann machine and Generative adversarial network. Future research opportunities and challenges of unsupervised techniques for medical image analysis have also been discussed.

**Keywords:** Unsupervised learning; medical image analysis; autoencoders; restricted Boltzmann machine; Deep belief network

## 1. Introduction

Medical imaging techniques, including magnetic resonance imaging (MRI), positron emission tomography (PET), computed tomography (CT), mammography, ultrasound, X-ray and digital pathology images, are frequently used diagnostic system for the early detection, diagnosis, and treatment of various complex diseases (Wani & Raza, 2018). In the clinics, the images are mostly interpreted by human experts such as radiologists and physicians. Because of major variations in pathology and the potential fatigue of human experts, scientists and doctors have started using computer-assisted interventions. The advancement in machine learning techniques, including deep learning, and availability of computing infrastructure through cloud computing, have given fuel to the field of computer-assisted medical image analysis and computer-assisted diagnosis (CAD). Deep learning is about learning representations, i.e, learning intermediate concept or features which are important to capture dependencies from input variables to output variables in supervised learning, or between subsets of variables in unsupervised learning. Both supervised and unsupervised machine learning approaches are widely applied in medical image analysis; each of them has their own pros and cons. Some of widely used supervised (deep) learning algorithms are Feedforward Neural Network (FFNN), Recurrent Neural Network (RNN), Convolutional Neural Network (CNN), Support Vector Machine (SVM) and so on (Jabeen et al., 2018). There are many scenarios where human supervisions are unavailable, inadequate or biased and therefore, supervised learning algorithm cannot be directly used. Unsupervised learning algorithms, including its deep architecture, give a big hope with lots of advantages and have been widely applied in several areas of medical and engineering problems including medical image analysis.

This chapter presents unsupervised deep learning models, its applications to medical image analysis, list of software tools/packages and benchmark datasets; and discusses opportunities and future challenges in the area.

## 2. Why Unsupervised Learning?

In the majority of machine learning projects, the workflow is designed in a supervised way, where the algorithm is guided by us what to do and what not to! In such supervised architecture the potential of the algorithms are limited in three ways, *(i)* A huge manual effort to create labels and *(ii)* Biases related to how it is being supervised, which probabilities the algorithm to think for other corner cases during problem solving, and *(iii)* Reduce the scalability of target function at hand.

To intelligently solve these issues, unsupervised machine learning algorithm can be used. Unsupervised machine learning algorithms not only derives insights directly from the data and group the data, but also uses these insights for data-driven decisions making. Also, unsupervised models are more robust in the sense that they act as a base for several different complex tasks where these can be utilized as the holy grail of learning and classification. In fact, the classification is not the only task that we do; rather, other tasks such as compression, dimensionality reduction, denoising, super resolution and some degree of decision making are also performed. Therefore, it is rather more useful to construct a model without knowing what tasks will be at hand and we will use representation (or model) for. In a nutshell, we can think of unsupervised learning as preparation (preprocessing) step for supervised learning tasks, where unsupervised learning of representation may allow better generalization of a classifier (Jabeen et al., 2018).

## 3. Taxonomy of Unsupervised Learning Tasks

In unsupervised learning, we group the unlabeled data set on the basis of underlying hidden features. By grouping data through unsupervised learning, at least we learn something about raw data.

### 3.1 Density estimation

Density estimation is one of the popular categories of unsupervised learning which discovers the intrinsic feature and structure of large and complex unlabeled data set via another non-probabilistic approach. Density estimation is a non-parametric method which doesn't possess much restriction and distributional assumption unlike parametric estimation.

Estimation of univariate or multivariate density function without any prior functional assumptions get almost limitless function from data. There are some widely used non-parametric methods of estimation.

*3.1.1 Kernel density estimation*

Kernel density estimation (KDE) uses statistical model to produce a probabilistic distribution that resembles an observed variable as a random variable. Basically, KDE is used for data smoothing, exploratory data analysis and visualization. A large number of kernels have been proposed, namely *normal Gaussian mixture model* and *multivariate Gaussian mixture model.* Some of the advantages of Kernel density estimation are:

- o No need for model specification (data speaks itself).
- o Estimation converges to any density, shape with sufficient sample.
- o Easily generalizes to higher dimension.
- o Densities are multivariate and multimodal with irregular cluster shape.

*3.1.2 Histogram density estimation*

Histogram based technique mainly adds smoothness of the density curve of reconstruction which can be optimized by kernel parameters and closely related to KNN density estimation algorithm (Bishop et al., 2006).

**3.2 Dimensionality reduction**

Why dimensionality reduction? There has been a tremendous increase in deployment of sensors and various imaging technique's which are being used in industry and medical diagnosis continuously record data and store it to be analyzed later. Lots of redundancy or noises are present initially when data are captured. For example, let us take a case of a patient having bone fracture. Initially doctors suggested for X-ray images which is a 2D/3D imaging, but when they do not find it helpful in diagnosis, then a CT scan and/or MRI (magnetic resonance imaging) may be taken which gives more detailed results for further right diagnosis. Now assume that an analyst sits with all this data to analyze and identified all important variables/dimensions which have most significant information's and left all unwanted parts of data. This is the problem of high unwanted dimension removal and needs treatment of dimension reduction. Dimension reduction is the process of reducing higher dimension data set into a lesser dimension, ensuring that final reduced data must convey equivalent information concisely.

Let's look at figure shown below. It shows two-dimensional x and y which are measurement of several objects in cm ($x_1$) and inches ($y_1$), if you continue to use both dimensions in machine learning problems it will introduce lots of noise in the system. So, it is better to just use one-dimension ($z_1$) and they will convey similar information.

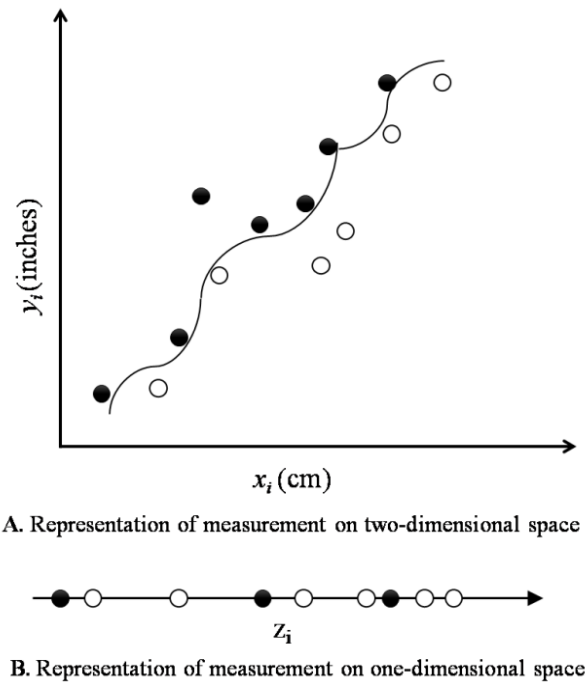

A. Representation of measurement on two-dimensional space

B. Representation of measurement on one-dimensional space

**Fig 1.** Representation of data in two dimensional and one dimensional space

There are some common methods to perform dimensionality reduction:

S.1.2    *Factor analysis*

Some variables in given data are highly correlated. These variables can be grouped on the basis of their correlations. This means a particular group can have highly correlated variable, but have low correlation with variables of other groups. Each group represents single inherent construct or factor. As compared to data having large number of dimensions, these factors are small in number, while these factors are difficult to find. There are two methods for doing factor analysis: *(i) Exploratory Factor Analysis, (ii) Confirmatory Factor Analysis*

*3.2.2 Principal component analysis*

A set of variables, which are linear combination of the original set of variables, performs higher dimensional space mapped to lower dimensions in such a way that variance of data in lower dimensional space is maximized. These new set of variables is known as principle components.

Let's consider a situation of two-dimensional data set, there can be only two principal components, first principal component is the most possible variation of original data and second principal component is orthogonal to the first principal component, as shown in Fig. 2.

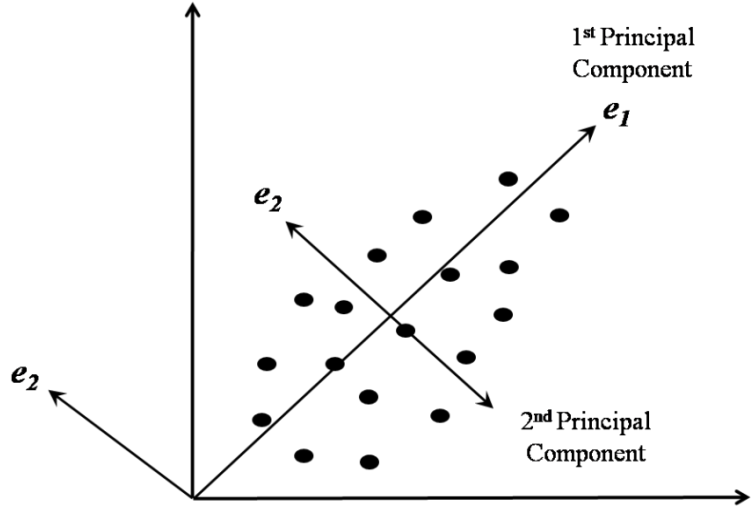

**Fig. 2** Principle components on a two-dimensional data

In practice, a simple principal component analysis (PCA) can be used to construct the covariance or correlation matrix of the data and compute the eigenvectors. The eigenvectors correspond to the largest eigenvalues (principal component) are used to reconstruct a large fraction of variance of original data. As a result, it is left with lesser number of eigenvector and original space has been reduced. There might be chances of loss of data, but it is retained by most important eigenvectors.

Consider a matrix $U(m)$ which stored empirical mean of input matrix R,

$$U(m) = \frac{1}{N}\sum_{n=1}^{N} R(m,n), \text{where } m = (1, 2, 3, \ldots\ldots M) \quad (1)$$

Calculate a normalized matrix $X$, $X = (R - U_e)$, where $e$ is a unitary vector matrix of size $N$. Finally, the mean square error ($E^2$) is calculated in which smallest eigenvalues are removed,

$$E^2 = trace(cov(x)) - \sum_{i=1}^{L} \lambda_i = \sum_{i=\lambda+1}^{M} \lambda_i \quad (2)$$

The *trace(A)* is the sum of all eigenvalues. Simple PCA is not capable of constructing nonlinear mapping, however, can implement nonlinear classification by using kernel techniques.

*3.2.3 Kernel PCA*

Kernel PCA is a nonlinear extension of conventional PCA, which is designed for dimensionality reduction of nonlinear subspaces depending on magnitude of input data and problem setup. In medical image analysis, hybrid kernel PCA is frequently used to get better results in unsupervised deep learning training model. Fischer et al. (2017) proposed an unsupervised deep learning illumination invariant kernel PCA, which is applied to each patch of respiratory signal extraction from X-ray fluoroscopy images leading to a set of low-dimensional embedding.

A kernel PCA comprised a kernel matrix K and kernel function $k(.)$ is a Mercer kernel (Minh et al. 2006), defined as $K_{ij} = k(x^{(i)}, x^{(j)})$, such that $k(.)$ return dot product of feature space. Now mapping of an eigenvalue of the kernel matrix, the Eigen decomposition and respected eigenvectors are computed as,

$$\lambda^{\{i\}} e^{\{i\}} = K\, e^{\{i\}} \tag{3}$$

$$D(x) = \frac{1}{e^{\{i\}}} \sum_{t=1}^{T} e^{\{i\}} k(x^{(t)}, x) \tag{4}$$

where $\lambda^{\{i\}}$ is eigenvalues and $e^{\{i\}}$ is eigenvectors of $K$; $T$ is the number of training sample $x$ to the principal component "$i$". Fischer et al. (2017) analyzed different methods like PCA, KPCA and Multi-Resolution PCA to Diaphragm tracking correlation coefficient between different versions of the same sequence and agreed that Multi-Resolution PCA produce the best result among most of the parameters. Principal component analysis network (PCANet) is a simple network architecture and one of the benchmark frameworks (Chan et al. 2015) for the unsupervised deep learning in recent time. However, Shi et al. (2017) propose an encoding as C-RBH-PCANet which is improved PCANet to effectively integrate the color pattern extraction and random binary hashing method for learning feature from color histopathological images.

## 3.3 Clustering

Clustering is an unsupervised classification of unlabeled data (patterns, data item or feature vectors) into similar groups (clusters) (Fig. 3). Cluster analysis is explanatory in nature to find structure in data (Jain, 2008). Some model of clustering includes semi-supervised clustering, ensemble clustering, simultaneous feature selection and large-scale data clustering were emerging as a hybrid-clustering. It involves analysis of multivariate data and applied in various scientific domains where clustering technique is utilized, such as machine learning, image analysis, bioinformatics, pattern recognition, computer vision and so on.

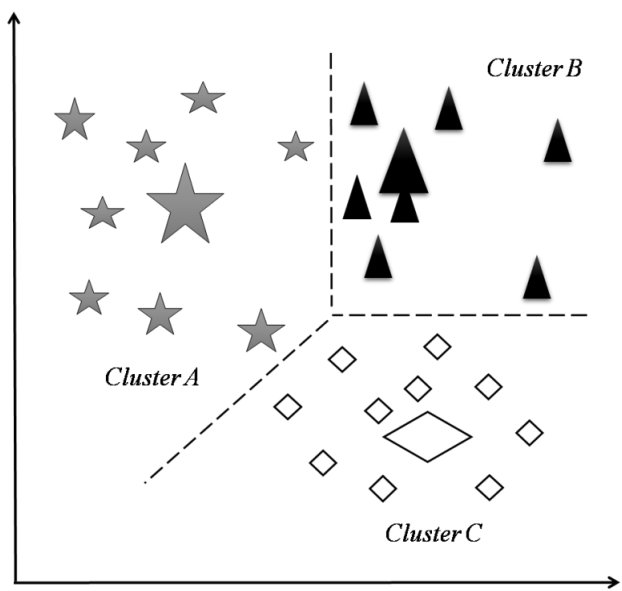

**Fig. 3** An illustration of data clustering

Clustering algorithm is broadly divided into two groups: *hierarchal clustering* and *partitional clustering*, as described below.

*3.3.1 Hierarchical clustering*

Hierarchical clustering algorithms find clusters recursively (using previously established cluster). These algorithms can be either in the agglomerative mode (bottom-up) in which begin with each element as a separate cluster, merge the most similar pair of clusters successively into large clusters, or in divisive (top-down) mode which begin with all elements in one cluster, recursively divide into smaller clusters. A hierarchical clustering algorithm yields a dendrogram representing group of patterns and similarity level (Jain et al., 1999). A detailed discussion can be found in Jain et al. (1999).

*3.3.2 Partitional (k-means) clustering*

One of the most popular partitioning clustering algorithms is k-means. In spite of several clustering algorithms published in over 50 years, k-means is still widely used (Jain, 2010). The most frequently used functions in partitional clustering is squared error criterion, which applied to isolate and compact clusters. Let $X = \{x_i: i = 1, 2, 3, .... N\}$ be the set of n d-dimensional elements clustered into set of K clusters as $C = \{c_k : k = 1, 2, 3,...K\}$. To find partitions, squared error between empirical mean of a cluster and elements in the cluster is minimized. Let $\mu_k$ be the mean of the cluster ($c_k$), the squared error between mean and elements in a cluster is defined as:

$$E(ck) = \sum_{xi \in ck} \|xi - \mu k\|^2 \qquad (5)$$

The main objective of K-means is to minimize the sum of squared error for all *k* clusters (Drineas et al., 1999).

$$E(c) = \sum_{k=1}^{k} \sum_{xi \in ck} \|xi - \mu k\|^2 \qquad (6)$$

Minimizing objective function is an NP-hard problem even for *k* = 2. Thus, k-means is a greedy algorithm and it can only be expected to converge to local minima.

## 4. Unsupervised deep learning models

This section introduces a formal introduction of unsupervised deep learning concepts, models and architectures that are used in medical image analysis. The unsupervised deep learning models can be roughly classified as shown in Fig. 4.

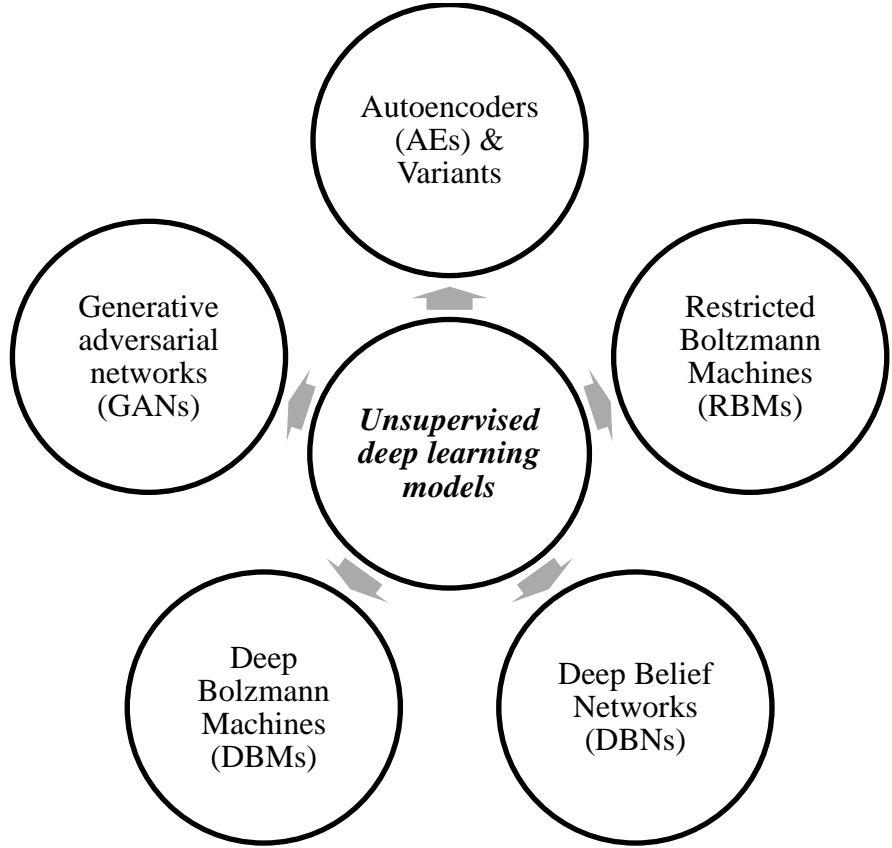

**Fig. 4** Unsupervised deep learning models

**4.1 Auto-encoders and its variants**

In the literature, autoencoders and its several variants are reported and are being extensively applied in medical image analysis.

*4.1.1 Autoencoders and Stacked autoencoder*

Autoencoders (AEs) (Bourlard et al., 1988) are simple unsupervised learning model consist single-layer neural network that transforms the input into a latent or compressed representation by minimizing the reconstruction errors between input and output values of the network. By constraining the dimension of latent representation (may be from different input) it is possible to discover relevant pattern from the data. AEs framework defines a feature to extract function with specific parameters (Bengio et al., 2013). Basically, AEs are trained with specific function $f_\theta$ is called encoder and $h = f_\theta(x)$ is feature vector or representation from input $x$, another parameterized function $g_\theta$ called decoder, producing input space back from feature space. In short, basic AEs are trained to minimize reconstruction error in finding a value of parameter $\theta$, given by,

$$T_{AE}(\theta) = \sum L\left(x, g_\theta[\![(f]\!]_\theta(x))\right) \qquad (7)$$

This minimization optionally followed by a non-linearity (most commonly used for encoder and decoder) as given by,

$$f_\theta(x) = S_f(W_x + b) \tag{8}$$

$$g_\theta(x) = S_g(W'_h + d) \tag{9}$$

where $S_f$ and $S_g$ are encoder and decoder activation function (normally, sigmoid, hyperbolic tangent or an identity function), respectively; parameters of model $\theta = \{W, b, W', d\}$, where $W$ and $W'$ are encoder decoder weight matrices, and $b$ and $d$ are encoder and decoder bias vector, respectively. Moreover, regularization or sparsity constraints may be applied in order to boost the discovery process. In case, hidden layer has the same input as the input layer, and no any non-linearity is added, the model would simply learn an identity function. Fig. 5(a) illustrates the basic structure of AE.

Stacked autoencoders (SAEs) are constructed by organizing AEs on top of each other also known as deep AEs. SAEs consist of multiple AEs stacked into multiple layers where the output of each layer is wired to the inputs of the successive layers Fig. 5(b). To obtain good parameters, SAE uses greedy layer-wise training. The benefit of SAE is that it can enjoy the benefits of deep network, which has greater expressive power. Furthermore, it usually captures a useful hierarchical grouping of the input (Shin et al., 2013).

*4.1.2 Denoising autoencoder*

Denoising autoencoder (DAEs) is another variant of the auto-encoder. Denoising investigated as a training criterion for learning to constitute better higher-level representation and extract useful features (Vincent et al. 2010). DAEs prevent the model from learning a trivial solution (Litjens G. et al., 2017) where the model is trained to reconstruct a clean input from the corrupted version from noise or another corruption. This is done by corrupting the initial input x into x̃ by using a stochastic function x̃ ~ $q_D$(x̃ |x). The corrupted input x̃ is then mapped to a hidden representation y = $f_\theta$(x̃) = s(W$_{x̃}$ + b) and reconstruct z = $g_{\theta'}$ (y). A schematic representation of DAE is shown in Fig.5(c). Parameter $\theta$ and $\theta'$ are initialized randomly and trained using stochastic gradient descent in order to minimize average reconstruction error. The denoising autoencoders continue minimizing same reconstruction loss between clean *X* and reconstruction from *Y*. This continues maximizing a lower bound on the mutual information between input x and representation y, and difference is obtained by applying mapping $f_\theta$ to a corrupted input. Hence, such learning is cleverer than the identity, and it extracts features useful for denoising.

Stack denoising autoencoder (SDAE) is a deep network utilizing the power of DAE (Bengio et al., 2007; Vincent et al., 2010) and RBMs in the deep belief network (Hinton & Salakhutdinov, 2006; Hinton et al., 2006).

*4.1.3 Sparse autoencoder*

The limitation of autoencoders to have only small numbers of hidden units can be overcome by adding a sparsity constraint, where a large number of hidden units can be introduced usually more than one input. The aim of sparse autoencoder (SAE) is to make a large number of neurons to have low average output so that neurons may be inactive most of the time.

Sparsity can be achieved by introducing a loss function during training or manually zeroing few strongest hidden unit activations. A schematic representation of SAE is shown in Fig. 5(d).

If the activation function of hidden neurons is $a_j$, the average activation function of each hidden neuron $j$ is given by

$$P_j = \frac{1}{m} \sum_{i=1}^{m} [a_j x_i] \qquad (10)$$

The objective of sparsity constraints is to minimize $P_j$ so that $P_j = P$, where $P$ is a sparsity constraint very close to 0 such as 0.05.

To enforce sparsity constraints, a penalty term is added to cost function which penalizes $\hat{P}_j$, de-weighting significantly from $P$. The penalty term is the Kullback-Leibler (KL) divergence between Bernoulli random variables, can be calculated as (Ng, 2013; Makhzani & Frey, 2013),

$$Penalty\ term = \sum_{j=1}^{N_2} KL(P||\hat{P}_j) \qquad (11)$$

where $N_2$ is number of neurons in the hidden layers, and index $j$ is summing over the hidden units in the network.

$$KL(P||\hat{P}_j) = P \log \frac{P}{\hat{P}_j} + (1-P) \log \frac{1-P}{1-\hat{P}_j} \qquad (12)$$

The property of penalty function is that $KL(P||\hat{P}_j) = 0$, if $P_j = \hat{P}_j$, otherwise it increases gradually as $\hat{P}_j$ diverses for $P$.

The k-sparse autoencoder (Makhzani & Frey 2013) is a form of sparse AE where k neurons having the highest activation function are chosen and the rest is ignored. The advantage of k-sparse AE is that they allow better exploration on a data set in terms of percentage activation of the network. The advantage of SAE is the sparsity constraints which penalize the cost function and as a result degrees of freedom is reduced. Hence, it regularizes and maintains the complexity of the network by preventing over-fitting.

*4.1.4 Convolutional autoencoder*

The most popular and widely used network model in deep unsupervised architecture is stacked AE. Stacked AE requires layer-wise pre-training. When layers go deeper during the pre-training process, it may be time consuming and tedious because of stacked AE is built with fully connected layers. Li et al. (2017) propose first trial to train convolutional directly an end-to-end manner without pre-training. Guo et al. (2017) suggested convolutional autoencoder (CAE) that is beneficial to learn feature for images and preserving the local structure of data and avoid distortion of feature space. A general architecture of CAE is depicted in Fig. 5(c).

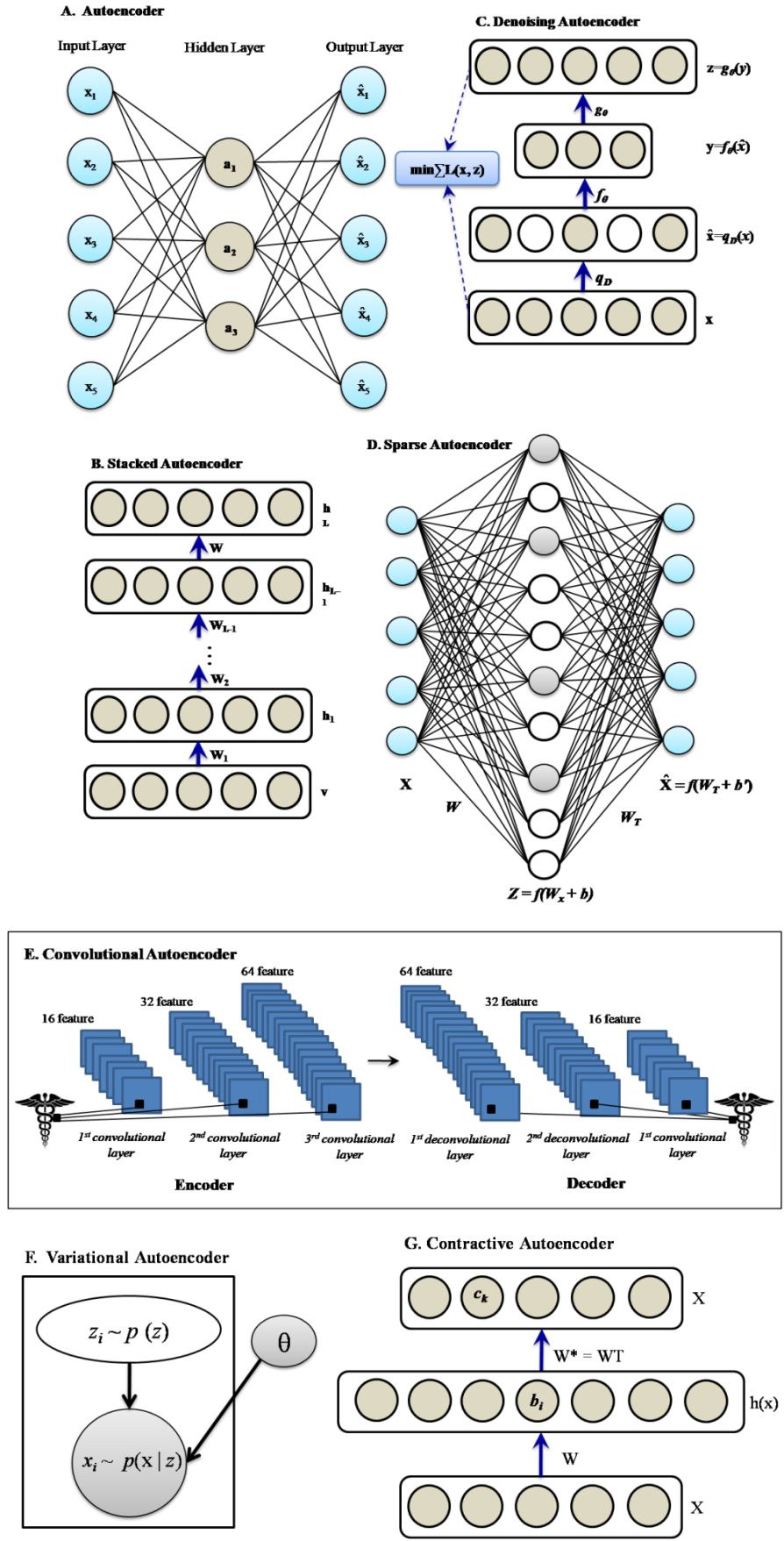

**Fig. 5 (a)-(g)** Diagrams showing networks of autoencoders and its different variants

*4.1.5 Variational autoencoder*

Another variant of autoencoder, called variational autoencoder (VAE), was introduced as a generative model (Kingma &Welling, 2013). A general architecture of VAE is given in Fig. 4(f). VAEs utilize the strategy of deriving a lower bond estimator from the directed graphical models with continuous distribution of latent variables. The generative parameter θ in the decoder (generative model) assist the learning process of the variational parameter, ϕ as encoder in the variational approximation model. VAEs apply the variational approach to latent representation, learning as additional loss component training estimators, known as stochastic gradient variational Bayes (SGVB) and Autoencoding variational Bayes (AEVB) (Kingma & Welling, 2013). It Optimizes the parameter ϕ and θ for probabilistic encoder $q_\phi(z|x)$, which is an approximation to the generative model $p_\theta(x, z)$, where z is latent variable and x is continuous or discrete variable. Its aim is to maximize the probability of each x in the training data set under entire generative process. However, alternative configuration of generative latent variable modeling rises to give deep generative models (DGMs) instead of existing assumption of symmetric Gaussian posterior (Partaourides at el., 2017).

*4.1.6 Contractive autoencoder*

Rifai (2011) presented a novel approach for training deterministic autoencoder. Contractive autoencoder is additional of explicit regularizer in the objective function that enables the model to learn a function having slight variations of input values. This additional regularizer corresponds to the squared Forbenius norm of the Jacobian matrix of given activation with respect to the input. The contractive autoencoder is obtained with the regularization term in following equation yield final objective function,

$$f_{CAE}(\theta) = \sum_{x \epsilon D_n} \left( L(x, g(f(x)) + \lambda \left\| J_f(x) \right\|_F^2 \right) \tag{13}$$

The difference between contractive AE and DAE stated by (Vincent et al., 2010) as contractive AE explicitly encourage robustness of representation, whereas DAE stressed on the robustness of reconstruction this property make sense of contractive AE a better choice than DAEs to learn useful feature extraction. Table 1 presents a summary of autoencoders and its variants, and Table 2 presents its applications for medical image analysis.

**Table 1.** Summary of autoencoders and its variants

| Types | Descriptions | References |
|---|---|---|
| Autoencoder | One of the simplest form which aims to learn a representation (encoding) for a set of data. | Ballard (1987); Bourlard & Kamp (1988) |
| Stacking autoencoder | An autoencoder having multiple layers where the outputs of each layers are given as inputs of the successive layer. | Zabalza et al. (2016) |
| Sparse autoencoder | Encourages hidden units to be zero or near to zero | Goodfellow et al. (2009) |
| Denosing autoencoder | Capable to predict true inputs from noisy data | LeCun & Gallinari, (1987); Vincent et al. (2008) |
| Convolutional autoencoder | Learn feature, preserve the local structure of data and avoid distortion of feature space | Guo et al. (2017) |
| Variational autoencoder | A generative model utilizing strategy of deriving a lower bond estimator from directed graphical models with continuous distribution of latent variables. | Kingma & Welling (2013) |
| Contractive autoencoder | Forces encoder to take small derivatives | Rifai et al. (2011) |

**Table 2** Applications of autoencoders and its variants for medical image analysis.
[Abbreviations: H&E: hematoxylin and eosin staining; AD: Alzheimer's disease; MCI: Mild cognitive impairment; fMRI: Functional magnetic resonance imaging; sMRI: Structural magnetic resonance imaging; rs-fMRI: Resting-state fMRI; DBN: Deep belief network; RBM: Restricted Boltzmann machine]

| Method | Task | Image type | Remarks | References |
|---|---|---|---|---|
| SAE | AD/MCI classification | MRI | SAE accompanied by supervised fine tuning | Suk & Shen (2013) |
| SAE | AD/MCI/HC classification | MRI & PET | Extraction of latent features on a huge set of features obtained from MRI and PET images using SAE | Suk et al. (2013a) |
| SAE | AD/MCI/HC classification | MRI | SAE used to pre-train 3D CNN | Payan & Montana (2015) |
| SAE | MCI/HC classification | fMRI | SAE used for feature extraction, HMM as a generative model on top | Suk et al. (2016) |
| SAE | Hippocampus segmentation | MRI | SAE used for representation learning and measure target/atlas patch | Guo et al. (2014) |
| SAE | Visual pathway segmentation | MRI | SAE used to learn appearance features to steer the shape model for segmentation | Mansoor et al. (2016) |
| SAE | Denoising DCE-MRI | MRI | Uses an ensemble of denoising SAE (pre-trained with RBMs). Denoising contrast-enhanced MRI sequences using expert DNNs (pre trained with RBMs) | Benou et al. (2016) |
| SSAE | Nucleus detection | Digital pathology image | detection of nuclei on breast cancer digital histopathological images. | Xu et al. (2016) |
| SAE | Stain normalization | Digital pathology image | SAE is applied to classify tissues and their subsequent histogram Matching | Janowczyk et al. (2017) |
| SAE | Density classification | Mammography | Unsupervised CNN with SAE to learn features from unlabeled data for breast texture and density classification | Kallenberg et al. (2016) |
| SAE | Lesion classification | MRI | Learn to extract features from multi-parametric MRI data, subsequently creates a hierarchical classification to detect prostate cancer. | Zhu et al. (2017) |
| SAE | Detection of Heart, kidney and liver location | MRI | SAE used for acquisition of spatio-temporal features on 2D along with time DCE-MRI | Shin et al. (2013) |
| SAE | Cell segmentation | Digital pathology image | Learning spatial relationships | Hatipoglu, N. 2017 |
| SAE | Segmentation right ventricle in cardiac MRI | MRI | SAE applied to obtain an initial right ventricle segmentation. | Avendi, M. 2017 |
| SDAE | Cell segmentation | Digital pathology image | The SDAE trained with data and their structured labels for cell segmentation | Su. H. at el 2018 |
| SSAE | AD | MRI | SSAE for early detection of Alzheimer's disease from brain MRI | Liu et al. (2014) |
| SDAE | Breast lesion | Ultrasound and CT | Stacked Denoising AE for Diagnosis of breast nodules and lesions | Cheng et al. (2016) |
| SDAE | Patient clinical events | Patient clinical history | SDAE for an unsupervised early prediction of patients e future clinical events and disease. | Miotto et al. (2016) |
| SDAE | -- | CT/MRI | Multi-modal SDAE used to pre-train the DNN. | Cheng et al. (2018) |
| DCAE | Modeling task fMRI | tfMRI | Deep Convolutional AE to model tfMRI. | Huang et al. (2018) |
| CAE | AD/MCI/HC classification | fMRI | CAE used to pre-train 3D CNN. | Hosseini-Asl et al. (2016) |
| CAE | Nucleus detection | Digital pathology image | Sparse CAE to detect and encode nuclei and feature extraction from tissue section images. | Hou et al. (2019) |

## 4.2. Restricted Boltzmann Machines

Restricted Boltzmann Machines (RBMs) are a variant of Markov Random Field (MRF), constitute of single layer undirected graphical model with an input layer or visible layer $x = (x_1, x_2...... x_N)$ and a hidden layer $h = \{h_1, h_2, .... H_M\}$. The connection between nodes/units are bidirectional, so each given input vector *x* can take the latent feature representation *h* and vice-versa. An RBM is a generative model which learns probability distribution over the given input space and generates new data point (Yoo, et al. 2014). Illustration of a typical RBM is shown in Fig. 6(a). In fact, RBMs are restricted version of Boltzmann machines where neurons must form an arrangement of bipartite graphs. Due to this restriction, pairs of nodes belonging to each of the visible and hidden nodes have a symmetric connection between them, and nodes within a group have no internal connections.. This restriction makes RBM more efficient training algorithm than the general case of Boltzmann machine. Hinton et al. (2010) proposed a practical guide to train RBMs.

RBMs have been utilized in various aspects of medical image analysis such as detection of variations in Alzheimer disease (Brosch, et al. 2013), image segmentation (Yoo et al. 2014), dimensionality reduction (Cheng et al. 2016), feature learning (Pereira et al. 2018) and so on. A brief account for the application of RMBs in medical image analysis is shown in Table 3.

**Table 3.** Applications of RBM for medical image analysis

| Method | Task | Image type | Remarks | References |
|---|---|---|---|---|
| RBM | AD | MRI | Uses a large dataset of MRI to rule out the mode of variations in AD brains. | Brosch et al. (2013) |
| RBM | Multiple sclerosis lesions | 3DMRI | Uses multi-channel 3D MR images of multiple sclerosis (MS) lesion for MS segmentation | Yoo et al. (2014) |
| RBM | AD/MCI/HC classification | MRI, PET | DBMs on multimodal images from MRI and PET scans for disease classification. | Suk et al. (2014) |
| RBM | Mass detection in breast cancer | Mammography | RBM based method for oversampling and semi-supervised learning to solve classification of imbalanced data with a few labeled samples | Cao et al. (2015) |
| RBM | fMRI blind source separation | fMRI | RBM used for both internal and functional interaction-induced latent source detection | Huang et al. (2016) |
| RBM | Vertebrae localization | CT, MRI | RBMs to locate the exact position of the vertebrae. | Cai et al. (2016b) |
| RBM | Benign/Malignant classification | Ultrasound | Shear wave elsatrography for class indication of benign and malignant mammary gland tumors using RBM. | Zhang et al. (2016a) |
| RBM | Tongue contour extraction | Ultrasound | Analysis of tongue motion during speech, using auto encoders in combination with RBM. | Jaumard-Hakoun et al. (2016) |
| CRBM | Lung tissue classification and airway detection | CT | Discriminative and generative learning by CRBM to develop filters for data training as well classification. | Van Tulder & de Bruijne (2016) |
| RBM | Cardiac arrhythmia classification | ECG | Achieves average recognition accuracy for ventricular and supraventricular ectopic beats (93.63% and 95.57%, respectively) for Cardiac arrhythmia classification. | Mathews et al. (2018) |
| RBM | Brain lesion segmentation | MRI | RBM is used for feature learning, and a Random Forest as a classifier. | Pereira et al. (2018) |

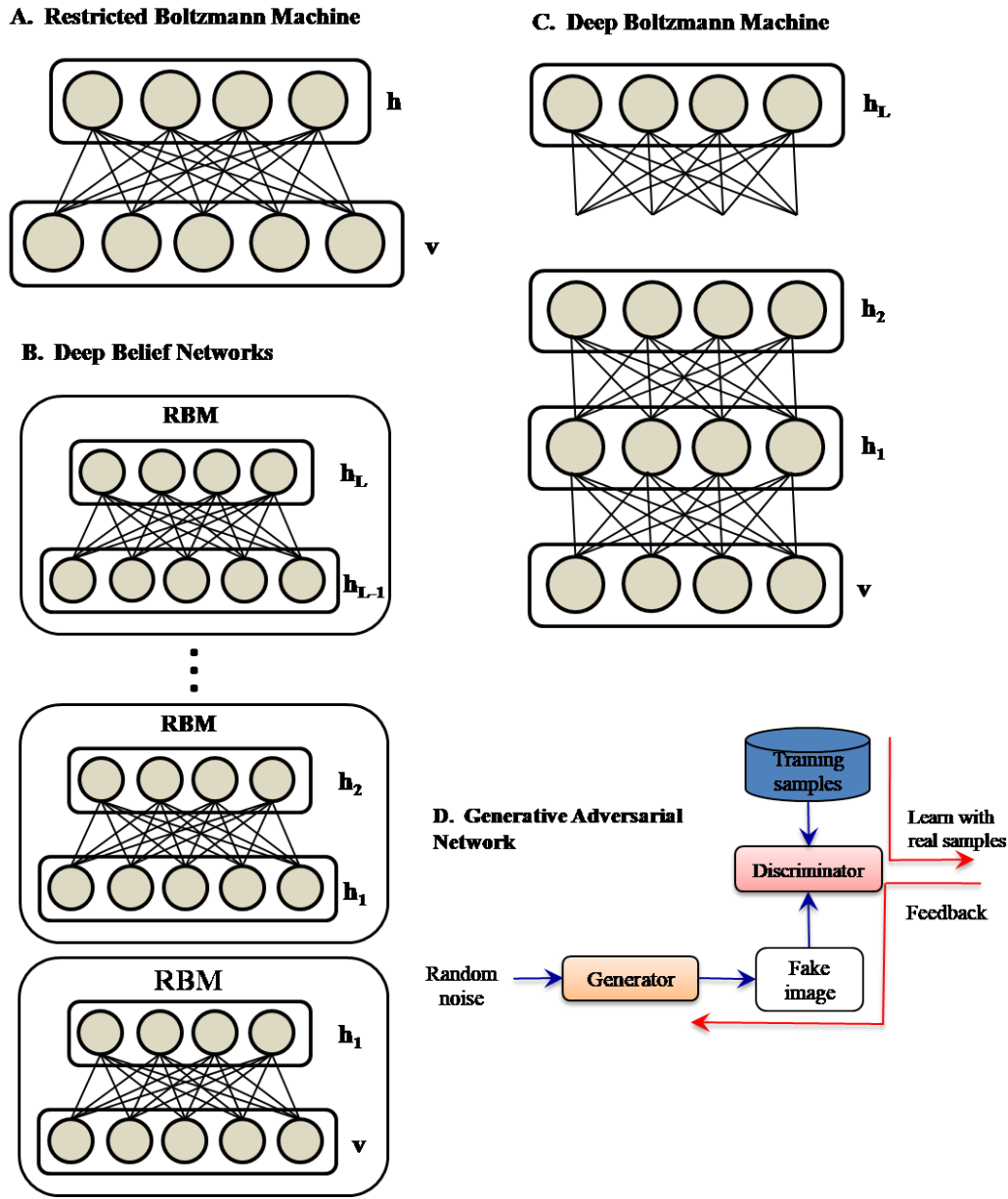

**Fig. 6 (a)-(d)** Diagrams showing various unsupervised network models

### 4.3. Deep Belief Networks

Deep Belief Networks (DBN) is a kind of neural network proposed by Bengio (2009). It is a greedy layer-wise unsupervised learning algorithm with several layers of hidden variables (Hinton et al., 2016). Layer-wise unsupervised training (Bengio 2007) help the optimization and weight initialization for better generalization. In fact, DBN is a hybrid single probabilistic generative model, like a typical RBM. In order to construct a deep architecture like SAEs where AEs layers are replaced by RBMs, DBN has one lowest visible layer **v**, representing state of input data vector and a series of hidden layers $h^1$, $h^2$, $h^3$, . . . $h^L$. When multiple RBMs are stacked hierarchically, an undirected generative model is formed by top two layers and directed generative model is formed by lower layers. Fig. 6(b), illustrates the structure of DBN. The following function in DBN represents the joint distribution of visible unit v, hidden layers $h^l$ (l = 1, 2…. L):

$$P(V, h^1, h^2 \ldots h^L) = \left(\prod_{l=0}^{L-2} P(h^l|h^{l+1})P(h^{L-1}, h^{(L)})\right) \qquad (14)$$

Hinton at el. (2006a) applied layer-wise training procedure, where lower layers learns low-level features and subsequently higher layers learns high-level features (Hinton at el. 1995). DBN are used to extract features from fMRI images (Plis et al., 2014), temporal ultrasound (Azizi et al. 2016), classify Autism spectrum disorders (Aghdam et al. 2018), and so on. Some of the applications of DBNs are presented in Table 4.

**Table 4.** Applications of DBNs for medical image analysis

| Method | Task | Image type | Remarks | References |
|---|---|---|---|---|
| DBN | AD/HC classification | MRI | DBNs with convolutional RBMs for manifold learning | Brosch & Tam (2013) |
| DBN, Convolutional RBM | Manifold Learning | MRI | DBM along with convolutional RBM layers to efficiently train DBMs in order to detect morphological changes in brain in normal as well as disease conditions | Brosch et al. (2014) |
| DBN | | MRI | Evaluation of DBN to estimate brain networks in neurocognitive disorders like Huntington's disease and Schizophrenia | Plis et al. (2014) |
| DBN | AD/MCI/HC classification | MRI | A group of voting schemes clubbed using an SVM to better classify AD and MCI from brain's 3D gray mater images. | Ortiz et al. (2016) |
| DBN | Left ventricle segmentation | Ultrasound | DBN assisted system exploiting non-rigid registration, landmarks and patches to maneuver multi atlas segmentation. | Carneiro et al. (2012); Carneiro & Nascimento (2013) |
| DBN | Schizophrenia/NH classification | MRI | Characterizing differences in morphology of various brain regions in schizophrenia using DBN and supervised fine tuning. | Pinaya et al. (2016) |
| DBN | Lesion classification | Ultrasound | Training DBN to extract features from prostate ultrasonography images to classify benign and malignant lesions. | Azizi et al. (2016) |
| DBN | Left ventricle segmentation | MRI | The combination of DBN and level set method to yield automated segmentation of the left ventricle from cardiac cine MRI | Ngo et al. (2017) |
| DBN | Cardiac arrhythmia classification | ECG | Achieves average recognition accuracy of ventricular and supraventricular ectopic beats (93.63% and 95.57%, respectively) for cardiac arrhythmia classification. | Mathews et al. (2018) |
| DBN | Autism spectrum disorders classification | rs-fMRI, sMRI | Classifies Autism spectrum disorders (ASDs) in children using rs-fMRI and sMRI data on the basis of Random Neural Network clustering. | Aghdam et al. (2018) |

## 4.4. Deep Boltzmann Machine

Deep Boltzmann machine (DBM) is a robust deep learning model proposed by Salakhutdinov et al. (2009) and Salakhutdinov et al. (2012). They stacked multiple RBMs in a hierarchal manner to handle ambiguous input robustly. Fig. 6(c) represents the architecture of DBM as a composite model of RBMs which clearly shows how DBM differ from DBN. Unlike DBNs, DBMs form undirected generative model combining information from both lower and upper layers which improves the representation power of DBMs. Training of layer-wise greedy algorithm for DBM (Salakhutdinov et al., 2015; Goodfellow et al., 2013b) is calculated by modifying in procedure of DBN.

Recently, a three-layer DBM was presented by Salakhutdinov et al. (2015) and Dinggang et al. (2017). In this three-layer DBM, to learn parameters $\theta = \{w^1, w^2\}$, the values of neighbour layer(s) and probability of visible and hidden units are computed using logistic sigmoidal function. The derivative of log likelihood of the observation ($V$) with respect to the model parameter ($\theta$) is computed as,

$$\frac{\partial}{\partial w(l)} \ln P(V;\theta) = E_{data}[h^{l-1}(h^l)^T] \sim E_{model}[h^{l-1}(h^l)^T] \qquad (15)$$

Where $E_{data}[.]$ denote data-dependent obtained from visible units and $E_{model}[.]$ denote data-independence obtained from the model. Some of the applications of DBMs are shown in Table 5.

**Table 5.** Applications of DBMs for medical image analysis

| Method | Task | Image type | Remarks | References |
|---|---|---|---|---|
| DBM | Heart motion tracking | MRI | Using three-layered Deep Boltzmann Machine to guide frame-by-frame heart segmentation during radiation therapy of cancer patient on cine MRI images. | Wu, et al., (2018) |
| DBN | AD/HC classification | MRI | DBN combined with convolutional RBMs for manifold learning. | Brosch & Tam (2013) |
| RBM | Breast-image classification | MRI | Restricted Boltzmann machine with backpropagation have been used for histopathological breast-image classification | Nahid et al., (2018) |
| DBM | Medical image retrieval | Multi digital image | DBM based multi model learning to learn joint density model. | Cao, et al., (2014) |

## 4.5. Generative Adversarial Network (GAN)

Generative Adversarial Network (GAN) (Goodfellow, et al. 2014) is one of recent promising technique for building flexible deep generative unsupervised architecture. Goodfellow et al. (2014) proposed two models generative model G and Discriminative model D, where G capture data distribution ($p_g$) over real data *t*, and *D* estimates the probability of a sample coming from training data (*m*) not from *G*. In every iteration, backpropagation generator and discriminator competing with each other. The training procedure the probability of D is maximized. This framework functions like a mini-max two-player game. The value function V(G, D) establishes following two-player mini-max game is given by,

$$\min_{G} \max_{D} V(G,D) = E_{t \sim p_{data}}[\log D(t)] + E_{m \sim p_m(m)}\left[\log\left(1 - D(G(m))\right)\right] \quad (16)$$

Where *D(t)* represents the probability of *t* from data *m* and $p_{data}$ is distribution of real-world data. This model seems to be stable and improved as $p_g = p_{data}$. A typical architecture of GAN is depicted in Fig. 6(d). In fact, these two adversaries, *Generator* and *Discriminator*, continuously battle during the processing of training. GAN have been applied to generate samples of photorealistic images to visualize new designs. Some of the applications of GAN for medical image analysis are presented in Table 6.

**Table 6.** Applications of GAN for medical image analysis

| Method | Task | Image type | Remarks | References |
|---|---|---|---|---|
| GAN | Synthesis of retinal images | Retinal images | MI-GAN generates precise segmented images for the application of supervised learning of retinal images. | Iqbal & Ali (2018) |
| GAN | Chest X-ray | X-ray | GAN used to produce photorealistic images which retain pathological quality | Canas, et al, (2018) |
| Dual GAN-FCN | Segmentation of regions of interest (ROIs) | --- | Improve GAN using dual-path adversarial learning for Fully Convolutional Network based image segmentation | Bi, et al., (2018) |
| GAN | Simulation of B-mode ultrasound images | Ultrasound | Conditional generative adversarial networks used to simulate ultrasound images at given 3D spatial locations. | Hu et al., (2017) |
| GAN | Treatment of lymphomas and lung cancer | PET | Multi-channel generative adversarial networks used to synthesize PET data. | Bi, et al., (2017) |

## 5. List of software tools/packages and benchmark datasets

A plethora of software tools and packages implementing unsupervised learning models (as discussed in the paper) has been developed and made available to the research community and data analysts. Some of the tools/packages and medical images benchmark datasets are listed in Table 7 and Table 8, respectively.

**Table 7.** List of software tools/packages for unsupervised learning models

| S. No. | Tools/ Packages Name | Models/ Methods | Description | Language /Technology | URL |
|---|---|---|---|---|---|
| 1. | `deeplearning4j` | Autoencoders | Deep learning APIs for Java having an implementation of several deep learning techniques. | Java | https://deeplearning4j.org/ |
| 2. | `unsup` under `torch7` | Autoencoder, etc. | A scientific computing framework with good support for machine learning algorithms that puts GPUs first. **Unsup** package provides few unsupervised learning algorithms such as autoencoders, clustering, etc. | Lua | https://github.com/torch/torch7 |
| 3. | `DeepPy` | Autoencoders | MIT licensed deep learning framework that runs on CPU or GPUs and implements autoencoders, in addition to other supervised learning algorithms. | Python | https://github.com/andersbll/deeppy http://andersbll.github.io/deeppy-website/ |
| 4. | `SAENET.train` | Stacked autoencoder | Build a stacked autoencoder in R environment for pre-training of feed-forward NN and dimension reduction of features. | R package | https://rdrr.io/cran/SAENET/man/SAENET.train.html |
| 5. | `kdsb17` | Convolutional autoencoder | Gaussian Mixture Convolutional Autoencoder (GMCAE) used for CT lung scan using Keras/TensorFlow | Python, Keras, Tensor-flow-gpu | https://github.com/alegonz/kdsb17 |
| 6. | `autoencoder` | Deep autoencoder | Training a deep autoencoder for MNIST digits datasets | Matlab | http://www.cs.toronto.edu/~hinton/code/Autoencoder_Code.tar |
| 7. | `H2O` | Deep autoencoder | Parallelized implementations of many supervised and unsupervised machine learning algorithms, including GLM, GBM, RF, DNN, K-Means, PCA, Deep AE, etc. | R package | https://cran.r-project.org/web/packages/h2o/ |
| 8. | `dbn` | DBN | Deep belief network pre-train in unsupervised manner with stacks of RBM, which in return fine-tuned DBN. | R package | https://rdrr.io/github/TimoMatzen/RBM/src/R/DBN.R |
| 9. | `darch` | DBN, RBM | Restricted Boltzmann machine, deep belief network implementation | R package | https://github.com/maddin79/darch |
| 10. | `deepnet` | DBN, RBM, deep autoencoders | Implementation of RBM, DBN, deep stacked autoencoders | R package | https://cran.r-project.org/web/packages/deepnet/ |
| 11. | `Vulpes` | DBN | DBN and other deep learning implementation in F#. | Visual Studio | https://github.com/fsprojects/Vulpes |
| 12. | `pydbm` | DBM/ RBM | RBM/DBM are implemented in python for pre-learning or dimension reduction | Python | https://pypi.org/project/pydbm/ |
| 13. | `RBM` | RBM | Simple RBM implementation in Python | Python | https://github.com/echen/restricted-boltzmann-machines |
| 14. | `xRBM` | RBM and its variants | Implementation of RBM and its variants in Tensorflow | Python | https://github.com/omimo/xRBM |
| 15. | `DCGAN.torch` | GAN | Unsupervised representation learning using Deep Convolutional GAN | Lua | https://github.com/soumith/dcgan.torch |
| 16. | `pix2pix` | GAN | Conditional Adversarial Networks for Image-to-image translation synthesizing from the image. | Linux Shell Script | https://github.com/phillipi/pix2pix |
| 17. | `ebgan` | GAN | Energy-based GAN equivalent to probabilistic GANs produces high resolution images. | Python | https://github.com/eriklindernoren/PyTorch-GAN/tree/master/implementations/ebgan |

**Table 8.** List of benchmark medical image datasets

[*Abbreviations.* ADNI: Alzheimer's Disease Neuroimaging Initiative; ABIDE: Autism Brain Imaging Data Exchange; DICOM: Digital Imaging and Communications in Medicine; BCDR: Breast Cancer Digital Repository; CIVM: Center for in Vivo Microscopy; DDSM: Digital Database for Screening Mammography; DRIVE: Digital Retinal Images for Vessel Extraction; IDA: Image & Data Archive; ISDIS: International Society for Digital Imaging of the Skin; NBIA: National Biomedical Imaging Archive; OASIS: Open Access Series of Imaging Studies; TCGA: The Cancer Genome Atlas; TCIA: The Cancer Imaging Archive]

| S. No. | Data set | Modalities | Medical condition | Accessibility | URL |
|---|---|---|---|---|---|
| 1. | ABIDE | MRI | Autism spectrum disorder | Open access | http://fcon_1000.projects.nitrc.org/indi/abide/ |
| 2. | ADNI | MRI | Alzheimer's disease | Paid | http://adni.loni.usc.edu/data-samples/access-data/ |
| 3. | BCDR | Mammography | Breast cancer | Open access | https://bcdr.eu/ |
| 4. | CIVM | 3D-MRM | Histology of the Embryonic and Neonatal Mouse | Limited access | http://www.civm.duhs.duke.edu/devatlas/ |
| 5. | DDSM | Mammography | Breast cancer | Open access | http://marathon.csee.usf.edu/Mammography/Database.html |
| 6. | DermNet | Photo dermatology | A huge database of various skin diseases | Limited access | http://www.dermnet.com/ |
| 7. | DICOM | MRI, CT, etc. | A variety of medical images, videos and signal files | Open access | https://www.dicomlibrary.com |
| 8. | DRIVE | 2D color images of retina | Retinal blood vessel segmentation to study diabetic retinopathy | Open access | http://www.isi.uu.nl/Research/Databases/DRIVE/download.php |
| 9. | IDA | | An online resource for neuroscience images | Open access | https://ida.loni.usc.edu/ |
| 10. | ISDIS | Dermoscopy, telemedicine, spectroscopy etc. | Skin disease | Paid | https://isdis.org/ |
| 11. | MedPix | Variety of imaging data | Online database of medical images, teaching cases, and clinical topics | Open access | https://medpix.nlm.nih.gov |
| 12. | NBIA | CT, PT, MRI, etc. | A database of the National Cancer Institute proving medical images of various conditions and anatomical sites. | Limited/ open access | https://imaging.nci.nih.gov/ |
| 13. | OASIS | MRI and PET | Normal aging or mild to moderate Alzheimer's Disease | Open access | http://www.oasis-brains.org/ |
| 14. | TCIA | Collection of MRI, CT etc. | Multimodal image archive for various types of cancer | Limited/ open access | http://www.cancerimagingarchive.net/ |
| 15. | TCGA | Histopathology slide images | Histopathology slide images from sample portions of various types of cancers | Open | https://cancergenome.nih.gov/ |

# 6. Discussion, opportunities and challenges

Medical imaging and diagnostic techniques are one of the most widely used for early detection, diagnosis and treatment of complex diseases. After significant advancement in machine learning and deep learning (both supervised and unsupervised), there is a paradigm shift from the manual interpretation of medical images by human experts such as radiologists and physicians to an automated analysis and interpretation, called computer-assisted diagnosis (CAD). As unsupervised learning algorithms can derive insights directly from data, use them for data-driven decisions making, and are more robust, hence they can be utilized as the holy grail of learning and classification problems. Furthermore, these models are also utilized for other important tasks including compression, dimensionality reduction, denoising, super resolution and some degree of decision making.

Unsupervised learning and CAD, both being in its infancy, researchers and practitioners have much opportunity in this area. Some of them are: *(i)* Allow us to perform exploratory analysis of data *(ii)* Allow to be used as preprocessing for supervised algorithm, when it is used to generate a new representation of data which ensure learning accuracy and reduces memory time overheads. *(iii)* Recent development of cloud computing, GPU-based computing, parallel computing and its cheaper cost allow big data processing, image analysis and execute complex deep learning algorithms very easily.

*Some of the challenges and research directions are:*

**(i) Difficult to evaluate whether algorithm has learned anything useful:** Due to lack of label in unsupervised learning, it is nearly impossible to quantify its accuracy. For instance, how can we access whether K-means algorithm found the right clusters? In this direction, there is a need to develop algorithms which can give an objective performance measure in unsupervised learning.

**(ii) Difficult to select right algorithm and hardware:** Selection of right algorithm for a particular type of medical image analysis is not a trivial task because performances of the algorithm are highly dependent on the types of data. Similarly, hardware requirement also varies from problem to problem.

**(iii) Will unsupervised learning work for me?** It is mostly asked question, but its answer totally depends on the problem at hand. In image segmentation problem, clustering algorithm will only work if the images do fit into naturals groups.

**(iv) Not a common choice for medical image analysis:** Unsupervised learning is not a common choice for medical image analysis. However, from literature it is revealed that these (autoencoders and its variants, DBN, RBM, etc.) are mostly used to learn the hierarchy level of features for classification tasks. It is expected that unsupervised learning will play pivotal role in solving complex medical imaging problems which are not only scalable to large amount of unlabeled data, but also suitable for performing unsupervised and supervised learning tasks simultaneously (Yi et al., 2018).

**(v) Development of patient-specific anatomical and organ model:** Anatomical skeletons play crucial role in understanding diseases and pathology. Patient-specific anatomical model is frequently used for surgery and interventions. They help to plan procedure, perform measurement for device surging, and predict the outcome of post-surgery complexities. Hence, the algorithm needs to be developed to construct patient-specific anatomical and organ model from medical images.

**(vi) Heterogeneous image data:** In the last two to three decades, more emphasis was given to well-defined medical image analysis applications, where developed algorithms were validated on well-defined types of images with well-defined acquisition protocol. The algorithms are required, which can work on more heterogeneous data.

**(vii) Semantic segmentation of images:** Semantic segmentation is task of complete scene understanding, leading to knowledge inference from imagery. Scene understanding is a core of computer vision problems which has several applications, including human-computer interaction, self-driving vehicles, virtual reality, and medical image analysis. The semantic segment of medical images with acceptable accuracy is still challenging.

**(viii) Medical video transmission:** Enabling 3D video in recently adopted telemedicine and U-healthcare applications result in more natural viewing conditions and better diagnosis. Also, remote surgery can be benefited from 3D video because of additional dimensions of depth. However, it is crucial to transmit data-hungry 3D medical video stream in real-time through limited bandwidth channels. Hence, efficient encoding and decoding techniques for 3D video data transmission is required.

**(ix) Need extensive inter-organizational collaborations:** Inter-professional and inter-organizational collaboration is important for better functioning of the health care system, eliminating some of the pitfalls such as limited resources, lack of expertise, aging populations, and combat chronic diseases (Karam et al., 2017). Medical image based CAD needs extensive inter-organizational collaborations among doctors, radiologists, medical image analysts, and computational data analysts.

**(x) Need to capitalize big medical imaging market:** According to IHS Markit report (https://technology.ihs.com.), medical imaging market has total global revenue of $21.2 billion in 2016, which is forecasted to touch $24.0 billion by 2020. According to WHO, global population will rise from 12% to 22% from 2015 to 2050. Population aging lead to increased rate of chronic diseases globally and hence there is a need to capitalize a big medical imaging market worldwide.

**(xi) Black-box and its acceptance by health professionals:** Machine learning algorithms are boon which solves the problems earlier thought to be unsolvable, however, it suffers from being "black-box", i.e., how output arrives from the model is very complicated to interpret. Particularly, deep learning models are almost non-interpretable and but still being used for complex medical image analysis. Hence, its acceptance by health professionals is still questionable.

**(xii) Will technology replace radiologists?** For the processing of medical images, deep learning algorithms help select and extract important features and construct new ones, leading to new representation of images, not seen before. For image interpretation side, deep learning helps identify, classify, quantify disease patterns, allow measure predictive targets, and make predictive models, and so on. So, will technology "replace radiologists", or migrate to "virtual radiologist assistant" in near future? Hence, following slogan is quite relevant in this context: "*Embrace it, it will make you stronger; reject it, it may make you irrelevant*".

In a nutshell, unsupervised learning is very much open topic where researchers can make contributions by developing a new unsupervised method to train how network (e.g. Solve a puzzle, generate image patterns, image patch comparison, etc.) and re-thinking of creating a great unsupervised feature representation, (e.g. What is the object and what is the background?), nearly analogous to the human visual system.

## 7. Conclusion

Medical imaging is one of the important techniques for early detection, diagnosis and treatment of complex diseases. Interpretation of medical images is usually performed by human experts such as radiologists and physicians. After the success of machine learning techniques, including deep learning, availability of cheap computing infrastructure through cloud computing, there has been a paradigm shift in the field of computer-assisted diagnosis (CAD). Both supervised and unsupervised machine learning approaches are widely applied in medical image analysis, each of them with their own pros and cons. Due to the fact that human supervisions are not always available or inadequate or biased, therefore, unsupervised learning algorithms, including its deep architecture, give a big hope with lots of advantages.

Unsupervised learning algorithms derive insights directly from data, and use them for data-driven decisions making. Unsupervised models are more robust and they can be utilized as the holy grail of learning and classification problems. These models are also used for other tasks including compression, dimensionality reduction, denoising, super resolution and some degree of decision making. Therefore, it is better to construct a model without knowing what tasks will be at hand and we would use representation (or model) for. In a nutshell, we can think of unsupervised learning as preparation (preprocessing) step for supervised learning tasks, where unsupervised learning of representation may allow better generalization of a classifier.

**Acknowledgements**

Authors would like to thank Ms. Sahar Qazi, Ms. Almas Jabeen, and Mr. Nisar Wani for necessary support.

**Conflict of Interest Statement**

Authors declare that there is no any conflict of interest in the publication of this manuscript.